\documentclass[aoas,MSNbibl,nameyear,dvips]{arximspdf}
\usepackage{amscd}
\usepackage{graphicx}

\doi{10.1214/13-AOAS653} 
\volume{7}
\issue{3}
\pubyear{2013}
\firstpage{1778}
\lastpage{1795}

\makeatletter

\renewcommand{\widehat}{\hat}

\newcommand{\var}{\operatorname{var}}
\newcommand{\tr}{\operatorname{tr}}

\newcommand{\real}{\mathbb{R}}

\makeatother

\begin{document}
\begin{frontmatter}

\title{Local adaptation and genetic effects on fitness: Calculations
for exponential family models with random effects}
\runtitle{Random effect aster models}

\begin{aug}
\author[A]{\fnms{Charles J.} \snm{Geyer}\corref{}\thanksref{m1}\ead[label=e1]{geyer@umn.edu}},
\author[B]{\fnms{Caroline E.} \snm{Ridley}\thanksref{m2}\ead[label=e2]{ridley.caroline@epa.gov}},
\author[C]{\fnms{Robert G.} \snm{Latta}\thanksref{m3}\ead[label=e3]{robert.latta@dal.ca}},\\
\author[D]{\fnms{Julie R.} \snm{Etterson}\thanksref{m4}\ead[label=e4]{jetterso@d.umn.edu}}
\and
\author[E]{\fnms{Ruth G.} \snm{Shaw}\thanksref{m1}\ead[label=e5]{shawx016@umn.edu}}
\runauthor{C. J. Geyer et al.}
\affiliation{University of Minnesota\thanksmark{m1},
US Environmental Protection Agency\thanksmark{m2},
Dalhousie University\thanksmark{m3} and
University of Minnesota, Duluth\thanksmark{m4}}
\address[A]{C. J. Geyer\\
School of Statistics\\
University of Minnesota\\
313 Ford Hall\\
224 Church St. SE \\
Minneapolis, Minnesota 55455\\
USA\\
\printead{e1}}
\address[B]{C. E. Ridley\\
US Environmental Protection Agency\\
1200 Pennsylvania Avenue, NW\\
William Jefferson Clinton\\
\quad Federal Building, Mail Code: 8623P\\
Washington, District of Columbia 20460\\
USA\\
\printead{e2}}
\address[C]{R. G. Latta\\
Department of Biology\\
Dalhousie University\\
Life Science Centre\\
1355 Oxford Street\\
PO BOX 15000\\
Halifax, Nova Scotia B3H 4R2\\
Canada\\
\printead{e3}}
\address[D]{J. R. Etterson\\
Department of Biology\\
University of Minnesota Duluth\hspace*{30.5pt}\\
153B Swenson Science Building\\
Duluth, Minnesota 55812\\
USA\\
\printead{e4}}
\address[E]{R. G. Shaw\\
Department of Ecology, Evolution\\
\quad and Behavior\\
University of Minnesota\\
100 Ecology\\
1987 Upper Buford Circle\\
St. Paul, Minnesota 55108\\
USA\\
\printead{e5}} 
\end{aug}

\received{\smonth{10} \syear{2012}}
\revised{\smonth{3} \syear{2013}}

%
\begin{abstract}
Random effects are implemented for aster models using two
approximations taken from Breslow and Clayton [\textit{J. Amer.
Statist. Assoc.} \textbf{88} (1993) 9--25]. Random effects
are analytically integrated out of the Laplace approximation to the
complete data log likelihood, giving a closed-form expression for an
approximate missing data log likelihood. Third and higher derivatives
of the complete data log likelihood with respect to the random effects
are ignored, giving a closed-form expression for second derivatives of
the approximate missing data log likelihood, hence approximate observed
Fisher information. This method is applicable to any exponential family
random effects model. It is implemented in the CRAN package
\texttt{aster} (R Core Team
[R: A Language and Environment for Statistical Computing
(\citeyear{rcore}) R Foundation for Statistical Computing],
Geyer [R package aster (2012)
\url{http://cran.r-project.org/package=aster}]). Applications are
analyses of local adaptation in the invasive California wild radish
(\emph{Raphanus sativus}) and the slender wild oat (\emph{Avena
barbata}) and of additive genetic variance for fitness in the partridge
pea (\emph{Chamaecrista fasciculata}).
\end{abstract}

%

\end{frontmatter}

\section{Introduction}

Aster models [\citet{gws,aster2}] are a partial generalization of generalized
linear models (GLM) that allow different components of the response vector
to have different families (some Bernoulli, some Poisson, some zero-truncated
Poisson, some normal) and also to be dependent, the dependence being specified
by a simple graphical model. Because of the way they incorporate dependence
among components of the
response, aster models are not GLM nor like other regression models
with which
statisticians are familiar, but they are special cases of graphical models
and of exponential families. Although aster models can be used whenever their
assumptions hold [for which see \citet{gws}], they were designed particularly
for life history analysis of plants and animals, which aims to model total
lifetime reproductive output (observed Darwinian fitness),
a random variable that fits no familiar
distribution, often having a large atom at zero (individuals that died before
producing offspring) as well as multiple modes.
Aster models can fit such data adequately by using data on other components
of fitness (survival in each year, number of flowers in each year,
number of seeds in each year, and number of seeds that germinate in
each year
for a plant and similar sorts of data for other organisms)
and modeling all these data jointly. It often turns out that, although
the marginal distribution of total lifetime reproductive output is intractable,
the conditional distribution of each component of the response vector given
some other component is tractable (e.g., number of seeds per flower
is Poisson). Biologists had recognized for decades that no statistical methods
before aster allowed valid statistical analysis of life history data
[\citet{aster2}], so aster models are becoming widely used.

Here we extend aster models to allow for random effects.
Our applications illustrate
three areas where random effects models are traditional. First,
when one categorical
predictor is nested within another, the effects for the nested predictor
are commonly treated as random, especially when they are nuisance parameters.
This is seen in both of our analyses
of local adaptation. Second, when levels of a categorical predictor
(such as
years) are not interesting in themselves but only as representatives
of a larger population, the corresponding effects are commonly treated
as random. This is seen in one of our analyses of local adaptation.
Third,
\citet{fisher18}, a paper that was the forerunner of all random effects models,
introduced the idea of random effects representing the cumulative
effects of
many genes. To obtain evolutionary predictions from life history analysis,
random effects models are necessary. This is seen in our analysis
of genetic variance for fitness. (Mapping the genes that contribute to
variation in fitness is not feasible; the number of them is so large,
and many are individually of such small effect, that it is unrealistic to
generate a sufficiently large study population to detect an informative
subset of them [\citet{travisano-shaw}]. If there were only a few genes
for fitness, then sequencing and ``machine learning'' would help,
but there is no sparsity here.)

As with GLM with random effects (generalized linear mixed models,
GLMM), aster models with random effects have analytically intractable
likelihoods necessitating the use of Monte Carlo, numerical integration
or approximate likelihood. Markov chain Monte Carlo likelihood
inference has a rich literature
[\citet
{penttinen,thompson,g+t,mcmcml,promislow,shaw-geyer-shaw,booth-hobert,hunter-et-al,okabayashi-geyer,hummel-et-al}],
but we have avoided it because it is very computationally intensive and
also very difficult for ordinary users to do correctly. Ordinary Monte
Carlo [\citet{sung}] has also been used, but is also very
computationally intensive. Numerical integration
[\citet{crouch-spiegelman}] is useful when there is only one variance
component but not otherwise [\citet{mccullough}, Section~7.2], and we
have avoided this too. Approximate integrated likelihood (AIL) is
based on the idea that if the complete data log likelihood were
quadratic in the random effects, then the random effects could be
integrated out analytically, and if the complete data log likelihood is
only close to quadratic in the random effects, then this is a
reasonable approximation, usually referred to as Laplace approximation
[\citet{b+c}]. For sufficiently large sample sizes and sufficiently few
random effects, the log likelihood is asymptotically expected to be
approximately quadratic [\citet{lecam-yang}, Chapter~6;
\citet{simple}], so this approximation may work well.

We use a second approximation, also introduced by \citet{b+c},
that is likewise an assumption that the log likelihood is close to quadratic
in the random effects. If the complete data log likelihood were exactly
quadratic in the random effects, then all derivatives higher than second
would be zero, and we assume this. Since the AIL already involves second
derivatives with respect to the random effects of the complete data log
likelihood, second derivatives of the log AIL would involve fourth derivatives
of the complete data log likelihood and would be computationally intractable.
This approximation allows us to compute approximate second derivatives
of the log AIL and hence approximate observed Fisher information.

\section{Theory of approximate integrated likelihoods} \label{secLAIL}

Although we are particularly interested in aster models, our
theory works for any exponential family model. The log likelihood can
be written
\[
l(\varphi) = y^T \varphi- c(\varphi),
\]
where $y$ is the canonical statistic vector,
$\varphi$ is the canonical parameter vector,
and the cumulant function $c$ satisfies
%
\begin{eqnarray}
\label{eqmoo}
\mu(\varphi) & = & E_{\varphi}(y) = c'(\varphi),
\\
\label{eqw}
W(\varphi) & = & \var_{\varphi}(y) = c''(\varphi),
\end{eqnarray}
where $c'(\varphi)$ denotes the vector of first partial derivatives
and $c''(\varphi)$ denotes the matrix of second partial derivatives.

We assume a canonical affine submodel with random effects determined by
%
\begin{equation}
\label{eqcan-aff-sub} \varphi= a + M \alpha+ Z b,
\end{equation}
where $a$ is a known vector, $M$ and $Z$ are known matrices, $b$
is a normal random vector with mean vector zero and
variance matrix $D$.
The vector $a$ is called the \emph{offset vector} and the matrices
$M$ and $Z$ are called the \emph{model matrices} for fixed and random
effects, respectively, in the terminology of the R function \texttt{glm}.
We assume the matrix $D$ is diagonal,
so the random effects are independent random variables.
The diagonal components of $D$ are called \emph{variance components}.

The unknown parameter vectors are $\alpha$ and $\nu$, where $\nu$ is
the vector of variance components. Thus, $D$ is a function of $\nu$,
although this is not indicated by the notation.
Typically each variance component corresponds to many random effects,
so each component of $\nu$ occurs multiple times as a diagonal element
of $D$.

In order to agree with the optimization literature, we prefer to minimize
rather than maximize. Thus, we use minus log likelihoods.
Minus the complete data log likelihood is
%
\begin{equation}
\label{eqfoo} - l(a + M \alpha+ Z b) + \tfrac{1}{2} b^T
D^{- 1} b + \tfrac{1}{2} \log\det(D)
\end{equation}
in case none of the variance components are zero.
We deal with the case of zero variance components in Sections~\ref{secroot}
and \ref{seczero}.

Let $b^*$ denote the result of minimizing (\ref{eqfoo}) considered
as a function of $b$ for fixed $\alpha$ and $\nu$.
Since minus the log likelihood of an exponential family is a convex function
[\citet{barndorff}, Theorem 9.1] and the middle term on the right-hand side
of (\ref{eqfoo}) is a strictly convex function, it follows that
(\ref{eqfoo}) considered as a function of $b$ for fixed $\alpha$ and
$\nu$
is a strictly convex function.
Moreover, this function has bounded level sets, because the first term
on the right-hand side of (\ref{eqfoo}) is bounded below
[\citet{gdor}, Theorems 4 and 6] and the second term has bounded level sets.
It follows that there is a unique global minimizer
[\citet{raw}, Theorems 1.9 and 2.6].
Thus, $b^*(\alpha, \nu)$ is well defined for all values
of $\alpha$ and $\nu$.

We define minus the log AIL to be
%
\begin{eqnarray}
\label{eqlog-pickle} q(\alpha, \nu) & = & - l\bigl(a + M \alpha+ Z b^*\bigr) +
\tfrac{1}{2} \bigl(b^*\bigr)^T D^{-1} b^*
\nonumber\\[-8pt]\\[-8pt]
&&{} + \tfrac{1}{2} \log \det \bigl[ Z^T W\bigl(a + M \alpha+ Z
b^*\bigr) Z D + I \bigr],\nonumber
\end{eqnarray}
where $I$ denotes the identity matrix of the appropriate dimension,
where $b^*$ is a function of $\alpha$ and $\nu$
and $D$ is a function of $\nu$, although this is not indicated by the
notation. Our equation (\ref{eqlog-pickle}) is the negation of
equation (5) in \citet{b+c}, who introduced the terminology
\emph{penalized quasi-likelihood} (PQL) for this approach.
Minimizing (\ref{eqlog-pickle}) gives approximate maximum likelihood
estimates of $\alpha$ and $\nu$, and
differentiating (\ref{eqlog-pickle}) twice gives an approximate observed
Fisher information matrix.

However, (\ref{eqlog-pickle}) is not easy to differentiate because $W$
is already the second derivative matrix of the cumulant function,
so second derivatives of (\ref{eqlog-pickle}) involve fourth derivatives
of the cumulant
function. For aster models there are no published
formulas for derivatives higher than second of the aster model cumulant
function and the software [the R package \texttt{aster},
\citet{package-aster}] does not compute them.
The derivatives do, of course, exist because every cumulant function of
a regular exponential family is infinitely differentiable at every
point of the canonical parameter space [\citet{barndorff}, Theorem 8.1].
Thus, we ignore derivatives higher than second, which is equivalent to
assuming $W$ is constant or that $c$ and $-l$ are quadratic.

This leads to the following idea. Rather than basing inference on
(\ref{eqlog-pickle}), we actually use
%
\begin{equation}
\label{eqlog-pickle-too} q(\alpha, \nu) = - l\bigl(a + M \alpha+ Z b^*\bigr) +
\tfrac{1}{2} \bigl(b^*\bigr)^T D^{-1} b^* +
\tfrac{1}{2} \log \det \bigl[ Z^T \widehat{W} Z D + I \bigr],\hspace*{-28pt}
\end{equation}
where $\widehat{W}$ is a constant matrix (not a function of $\alpha$ and
$\nu$). This makes sense for any
choice of $\widehat{W}$ that is symmetric and positive semidefinite,
but we will choose $\widehat{W}$ that are close to
$W(a + M \hat{\alpha} + Z \hat{b})$, where
$\hat{\alpha}$ and $\hat{\nu}$ are the joint minimizers
of (\ref{eqlog-pickle}) and $\hat{b} = b^*(\hat{\alpha}, \hat{\nu})$.
Note that (\ref{eqlog-pickle-too}) is a redefinition of $q(\alpha,
\nu)$.
Hereafter we will no longer use the definition (\ref{eqlog-pickle}).

Introduce
%
\begin{equation}
\label{eqkey} p(\alpha, b, \nu) = - l(a + M \alpha+ Z b) + \tfrac{1}{2}
b^T D^{-1} b + \tfrac{1}{2} \log \det \bigl[
Z^T \widehat{W} Z D + I \bigr],\hspace*{-22pt}
\end{equation}
where, as the left-hand side says, $\alpha$, $b$ and $\nu$ are all
free variables and, as usual, $D$ is a function of $\nu$.
Since the terms that contain $b$ are the same in both (\ref{eqfoo})
and (\ref{eqkey}), $b^*$ can also be defined as the result of
minimizing (\ref{eqkey}) considered
as a function of $b$ for fixed $\alpha$ and $\nu$.
Thus, (\ref{eqlog-pickle-too}) is a profile of (\ref{eqkey})
and $(\hat{\alpha}, \hat{b}, \hat{\nu})$ is the joint minimizer
of (\ref{eqkey}).

We now switch notation for partial derivatives, using subscripts to
indicate derivatives,
explained in more detail in
Section~1.6 of the accompanying technical report [\citet{aster-random}].
Then second derivatives of (\ref{eqlog-pickle-too}) can be written
using the implicit function theorem and the fact that $b^*$ minimizes
(\ref{eqkey}) as
\begin{eqnarray*}
q_{\alpha\alpha}(\alpha, \nu) & = & p_{\alpha\alpha}\bigl(\alpha, b^*, \nu\bigr) -
p_{\alpha b}\bigl(\alpha, b^*, \nu\bigr) p_{b b}\bigl(\alpha, b^*,
\nu\bigr)^{-1} p_{b \alpha}\bigl(\alpha, b^*, \nu\bigr),
\\
q_{\alpha\nu}(\alpha, \nu) & = & p_{\alpha\nu}\bigl(\alpha, b^*, \nu\bigr) -
p_{\alpha b}\bigl(\alpha, b^*, \nu\bigr) p_{b b}\bigl(\alpha, b^*,
\nu\bigr)^{-1} p_{b \nu}\bigl(\alpha, b^*, \nu\bigr),
\\
q_{\nu\nu}(\alpha, \nu) & = & p_{\nu\nu}\bigl(\alpha, b^*, \nu\bigr) -
p_{\nu b}\bigl(\alpha, b^*, \nu\bigr) p_{b b}\bigl(\alpha, b^*,
\nu\bigr)^{-1} p_{b \nu}\bigl(\alpha, b^*, \nu\bigr),
\end{eqnarray*}
a particularly simple and symmetric form
[for a detailed derivation see Sections~1.7 and 1.8 of \citet{aster-random}].
If we combine all the parameters
in one vector $\psi= (\alpha, \nu)$ and write $p(\psi, b)$ instead
of $p(\alpha, b, \nu)$, we have
%
\begin{equation}
\label{eqpsi-psi} q_{\psi\psi}(\psi) = p_{\psi\psi}\bigl(\psi, b^*\bigr)
- p_{\psi b} \bigl(\psi, b^* \bigr) p_{b b} \bigl(\psi, b^*
\bigr)^{- 1} p_{b \psi} \bigl(\psi, b^* \bigr).
\end{equation}
This form is familiar from the conditional variance formula
for normal distributions; if
%
\begin{equation}
\label{eqfat} \pmatrix{ \Sigma_{1 1} & \Sigma_{1 2}
\cr
\Sigma_{2 1} & \Sigma_{2 2} }
\end{equation}
is the partitioned variance matrix of a partitioned normal random vector
with components $X_1$ and $X_2$, then the variance matrix of the conditional
distribution of $X_1$ given $X_2$ is
%
\begin{equation}
\label{eqthin} \Sigma_{1 1} - \Sigma_{1 2}
\Sigma_{2 2}^{- 1} \Sigma_{2 1},
\end{equation}
assuming that $X_2$ is nondegenerate [\citet{anderson}, Theorem 2.5.1].
Moreover, if the conditional distribution is degenerate, that is, if there
exists a nonrandom vector $v$ such that $\var(v^T X_1 \mid X_2) = 0$, then
\[
v^T X_1 = v^T \Sigma_{1 2}
\Sigma_{2 2}^{- 1} X_2
\]
almost surely, assuming $X_1$ and $X_2$ have mean zero
[also by \citet{anderson}, Theorem 2.5.1], and the joint distribution
of $X_1$ and $X_2$ is also degenerate. Thus, we conclude that if
the (joint) Hessian matrix of $p$ is nonsingular, then so is the (joint)
Hessian matrix of $q$ given by (\ref{eqpsi-psi}).

The second derivatives of $p$ we need for the second derivatives of $q$ are
\begin{eqnarray*}
p_{\alpha\alpha}(\alpha, b, \nu) & = & M^T W(a + M \alpha+ Z b) M,
\\
p_{\alpha b}(\alpha, b, \nu) & = & M^T W(a + M \alpha+ Z b) Z,
\\
p_{b b}(\alpha, b, \nu) & = & Z^T W(a + M \alpha+ Z b) Z +
D^{- 1},
\\
p_{\alpha\nu_k}(\alpha, b, \nu) & = & 0,
\\
p_{b \nu_k}(\alpha, b, \nu) & = & - D^{- 1} E_k
D^{- 1} b,
\\
p_{\nu_j \nu_k}(\alpha, b, \nu) & = & b^T D^{- 1}
E_j D^{- 1} E_k D^{- 1} b
\\
&&{} - \tfrac{1}{2} \tr \bigl( \bigl[ Z^T \widehat{W} Z D + I
\bigr]^{- 1} Z^T \widehat{W} Z E_j
\\
&&
\hspace*{29.7pt}{}\times\bigl[ Z^T \widehat{W} Z D + I \bigr]^{- 1}
Z^T \widehat{W} Z E_k \bigr),
\end{eqnarray*}
where $E_k = D_{\nu_k}$
[for a detailed derivation see Section~1.8 of \citet{aster-random}].
In our use of the implicit function theorem we needed
$p_{b b}(\alpha, b^*, \nu)$ to be invertible. From the explicit form
given above we see that it is actually positive definite, because
$W(a + M \alpha+ Z b)$ is positive semidefinite by (\ref{eqw}).

\section{Square roots of variance components} \label{secroot}

It is part of the folklore
of random effects models that introducing
square roots of variance components avoids issues with zero variance
components and with constrained optimization.
Introduce new parameters by $\nu_j = \sigma_j^2$
and new random effects by $b = A c$,
where $A$ is diagonal\vadjust{\goodbreak} and $A^2 = D$.
Then the objective function (\ref{eqkey}) becomes
%
\begin{equation}
\label{eqkey-rooted}\quad \tilde{p}(\alpha, c, \sigma) = - l(a + M \alpha+ Z A c) +
\tfrac{1}{2} c^T c + \tfrac{1}{2} \log \det \bigl[
Z^T \widehat{W} Z A^2 + I \bigr].
\end{equation}
There are now no constraints (the $\sigma_j$ are allowed to be negative)
and (\ref{eqkey-rooted}) is a continuous function of all variables
(there is no discontinuity when $\sigma_j = 0$).

We find this change-of-parameter useful and use it to avoid constrained
optimization [R package \texttt{aster}, \citet{package-aster}].
However, it also causes problems.

First, it introduces spurious zeros of the
first derivative of (\ref{eqkey-rooted}) that are not stationary
points of
(\ref{eqkey}). In fact, the partial of (\ref{eqkey-rooted}) with respect
to $\sigma_j$ is always zero when $\sigma_j = 0$ by symmetry. Thus, first
derivatives of (\ref{eqkey-rooted}) cannot be used to test whether the
minimum occurs when some variance component is zero. Since the issue of
whether a variance component is zero is often of scientific interest,
this is very problematic. We solve this problem by looking at
first derivatives of (\ref{eqlog-pickle-too}) on the original
parameter scale
(Section~\ref{seczero} below) and using the theory of constrained
optimization.

Second, the formula (\ref{eqpsi-psi}) for observed Fisher information,
although guaranteed to be positive definite if infinite precision arithmetic
is used, is not so guaranteed if it is evaluated by
the usual computer arithmetic (with 16
decimal digit precision). We found that the analog of
(\ref{eqpsi-psi}) after the change of parameter from $\nu$ to
$\sigma$ was
even more computationally unstable.

Thus, although (\ref{eqkey-rooted}) is useful for finding approximate
maximum likelihood estimates, we find it problematic for calculating approximate
observed Fisher information or for determining whether approximate maximum
likelihood estimates of variance components are zero.

\section{Theory of constrained optimization} \label{seczero}

In order to determine whether the minimizer of (\ref{eqkey}) occurs
on the
boundary of the parameter space where some variance component is zero, we
need to use the theory of constrained optimization.
Unfortunately, we cannot use the Karush--Kuhn--Tucker theory
[\citet{fletcher}, Section~9.1; \citet{nocedal-wright}, Section~12.2],
which is familiar to some statisticians, because the constraint set is not
determined by smooth inequality constraints. More advanced nonsmooth
analysis [\citet{raw}] does handle our problem, but is unfamiliar to most
statisticians. Fortunately, for our analysis
we can use a simplification of the latter
theory based on the notion of directional derivatives.
The technical report [\citet{aster-random}] uses the full theory from
\citet{raw}, but the results are the same as those stated here in
terms of directional derivatives.

\subsection{Incorporating constraints in the objective function}

The formula (\ref{eqkey}) makes sense when all variance components are
positive (so $D$ is invertible). Otherwise, it does not.
As is common in nonsmooth analysis [\citet{raw}, Section 1A],
we define the objective function to have the value $+ \infty$ off
of the constraint set. Since $+ \infty$ can never minimize the objective
function, this incorporates the constraints in the objective function.
On the boundary of the constraint set (where some variance
components are zero and the corresponding random effects are also zero)
we extend the objective function by lower semicontinuity.

Since all but the middle term on the right-hand side of
(\ref{eqkey}) are actually defined on some neighborhood of each point
of the constraint set and differentiable at each point
of the constraint set, we only need to deal with the middle term.
Define
%
\begin{equation}
\label{eqh-lsc} h(b, \nu) = \cases{b^2 / \nu, &\quad $\nu> 0$,
\cr
0, &\quad
$\nu= 0$ and $b = 0$,
\cr
+ \infty, &\quad otherwise.}
\end{equation}
Let $\nu_{k(i)}$ denote the variance of $b_i$, and
let $\dim(b)$ denote the number of random effects. Then (\ref{eqkey})
can be rewritten
%
\begin{eqnarray}
\label{eqkey-lsc} p(\alpha, b, \nu) & = & - l(a + M \alpha+ Z b) +
\frac{1}{2} \sum_{i = 1}^{\dim(b)}
h(b_i, \nu_{k(i)})
\nonumber\\[-8pt]\\[-8pt]
&&{} + \frac{1}{2} \log \det \bigl[ Z^T \widehat{W} Z D + I
\bigr],\nonumber
\end{eqnarray}
where $h$ is given by (\ref{eqh-lsc}), provided all of the components of
$\nu$ are nonnegative. The proviso is necessary because the third term
on the right-hand side is not defined for all values of $\nu$, only those
such that the argument of the determinant is a positive definite matrix.
Hence, we must separately define $p(\alpha, b, \nu) = + \infty$
whenever any component of $\nu$ is negative.

\subsection{Directional derivatives}

A necessary condition for a local minimum of a smooth function is that
the first derivative is zero (Fermat's rule).
This works at points in the interior of the constraint set where
(\ref{eqkey-lsc}) is differentiable. It does not work at points on
the boundary.
There we need what Rockafellar and Wets [(\citeyear{raw}), Theorem 10.1]
call \emph{Fermat's rule, generalized:} a necessary condition for a local
minimum is that all directional derivatives are nonnegative.

For any extended-real-valued function $f$ on $\real^d$,
the directional derivative of $f$ at the point $x$ in the direction $w$
is defined by
\[
d f(x) (w) = \lim_{\tau\searrow0} \frac{f(x + \tau w) - f(x)}{\tau}.
\]
At a point $x$ where $f$ is differentiable, we have
$d f(x)(w) = w^T f'(x)$, and the notion of directional derivatives gives
no information that cannot be obtained from partial derivatives. It is
only on the boundary where we need directional derivatives.

In the interior of the constraint set, where this function is smooth,
ordinary calculus gives
\[
d h(b, \nu) (u, v) = \frac{2 b u}{\nu} - \frac{b^2 v}{\nu^2},
\]
where the notation on the left-hand side means the directional
derivative of $h$
at the point $(b, \nu)$ in the direction $(u, v)$.
On the boundary of the constraint set, which consists of the single point
$(0, 0)$, the directional derivatives are given by
\[
d h(0, 0) (u, v) = h(u, v).
\]

\subsection{Applying the generalization of Fermat's rule}

This theory tells us nothing we did not already know about points
in the interior of the constraint set. The only way we can
have $d f(x)(w) \ge0$ for all vectors $w$ is if $f'(x) = 0$.
It is only at points on the boundary of the constraint set, where
directional derivatives are the key.

Even on the boundary, the conclusions of the theory about components
of the state that are not on the boundary agree with what we already knew.
At a local minimum we have
%
\begin{equation}
\label{eqdescent-alpha} p_\alpha(\alpha, b, \nu) = 0
\end{equation}
and
%
\begin{eqnarray}
\label{eqdescent-other-smooth}
p_{\nu_j}(\alpha, b, \nu) & = & 0,\qquad \mbox{$j$
such that $\nu_j > 0$},
\nonumber\\[-8pt]\\[-8pt]
p_{b_i}(\alpha, b, \nu) & = & 0,\qquad \mbox{$i$ such that $\nu _{k(i)}
> 0$}\nonumber
\end{eqnarray}
[\citet{aster-random}, Section~1.10.4, gives details].

Thus, assuming that we are at a point $(\alpha, b, \nu$)
where (\ref{eqdescent-alpha})
and (\ref{eqdescent-other-smooth}) hold,
and we do assume this throughout the rest of this section,
$d p(\alpha, b, \nu)(s, u, v)$ actually involves only $v_j$ and $u_i$ such
that $\nu_j = 0$ and $k(i) = j$. Define
%
\begin{equation}
\label{eqkey-smooth} \bar{p}(\alpha, b, \nu) = - l(a + M \alpha+ Z b) +
\tfrac{1}{2} \log \det \bigl[ Z^T \widehat{W} Z D + I \bigr]
\end{equation}
[the part of (\ref{eqkey-lsc}) consisting of the smooth terms].
Then
%
\begin{eqnarray}
\label{eqdescent-other-nonsmooth}
&& d p(\alpha, b, \nu) (s, u, v) \nonumber\\[-8pt]\\[-8pt]
&&\qquad = \sum
_{j \in J} \biggl[ v_j \bar{p}_{\nu_j}(
\alpha, b, \nu)
+ \sum_{i \in k^{- 1}(j)} \bigl( u_i
\bar{p}_{b_i}(\alpha, b, \nu) + h(u_i, v_j)
\bigr) \biggr],\nonumber
\end{eqnarray}
where $J$ is the set of $j$ such that $\nu_j = 0$,
where $k^{- 1}(j)$ denotes the set of $i$ such that $k(i) = j$,
and where $h$ is defined by (\ref{eqh-lsc}).
To check that we are at a local minimum, we need to show
that (\ref{eqdescent-other-nonsmooth}) is nonnegative
for all vectors $u$ and $v$.
Conversely, to verify that we are not at a local minimum, we need to find
one pair of vectors $u$ and $v$ such that (\ref{eqdescent-other-nonsmooth})
is negative. Such a pair $(u, v)$ we call a \emph{descent direction}.
Since Fermat's rule generalized is a necessary but not sufficient condition
(like the ordinary Fermat's rule), the check that we are at a local minimum
is not definitive, but the check that we are not is. If a descent direction
is found, then moving in that direction away from
the current value of $(\alpha, b, \nu)$ will decrease the objective
function (\ref{eqkey-lsc}).

So how do we find a descent direction? We want to minimize
(\ref{eqdescent-other-nonsmooth}) considered as a function of $u$ and $v$
for fixed $\alpha$, $b$ and $\nu$.
We can consider the terms of
(\ref{eqdescent-other-nonsmooth}) for each $j$ separately.
If the minimum of
%
\begin{equation}
\label{eqdescent-other-nonsmooth-j} v_j \bar{p}_{\nu_j}(
\alpha, b, \nu) + \sum_{i \in k^{- 1}(j)} \bigl( u_i
\bar{p}_{b_i}(\alpha, b, \nu) + h(u_i, v_j)
\bigr)
\end{equation}
over all vectors $u$ and $v$ is nonnegative, then the minimum is zero,
because (\ref{eqdescent-other-nonsmooth-j}) has the value zero
when $u = 0$ and $v = 0$. Thus, we can ignore this $j$ in calculating
the descent direction.

Since we are only interested in finding a descent direction, the length
of the direction vector does not matter.
Thus, we can
do a constrained minimization of~(\ref{eqdescent-other-nonsmooth-j}),
constraining $(u, v)$ to lie in a ball. This is found by the well-known
Karush--Kuhn--Tucker theory of constrained optimization
[\citet{fletcher}, Section~9.1; \citet{nocedal-wright}, Section~12.2]
to be the minimum of the Lagrangian function
%
\begin{equation}
\label{eqlaggard}\qquad L(u, v) = \lambda v_j^2 +
v_j \bar{p}_{\nu_j}(\alpha, b, \nu) + \sum
_{i \in k^{- 1}(j)} \biggl( \lambda u_i^2 +
u_i \bar{p}_{b_i}(\alpha, b, \nu) + \frac{u_i^2}{v_j}
\biggr),
\end{equation}
where $\lambda> 0$ is the Lagrange multiplier, which would have to be
adjusted if we were interested in constraining $(u, v)$ to lie in a particular
ball. Since we do not care about the length of $(u, v)$, we can use any
$\lambda$. We have replaced $h(u_i, v_i)$ by $u_i^2 / v_j$ because we
know that if we are finding an actual descent direction, then we will
have $v_j > 0$. Now
\begin{eqnarray*}
L_{u_i}(u, v) & = & 2 \lambda u_i + \bar{p}_{b_i}(
\alpha, b, \nu) + \frac{2 u_i}{v_j},\qquad i \in k^{- 1}(j),
\\
L_{v_j}(u, v) & = & 2 \lambda v_j + \bar{p}_{\nu_j}(
\alpha, b, \nu) - \sum_{i \in k^{- 1}(j)} \frac{u_i^2}{v_j^2}.
\end{eqnarray*}
The minimum occurs where these are zero.
Setting the first equal to zero and solving for $u_i$ gives
\[
\hat{u}_i(v_j) = - \frac{\bar{p}_{b_i}(\alpha, b, \nu)}{2 (\lambda+ 1 / v_j)},
\]
plugging this back into the second gives
\[
L_{v_j} \bigl(\hat{u}(v), v \bigr) = 2 \lambda v_j +
\bar{p}_{\nu_j}(\alpha, b, \nu) - \frac{1}{4 (\lambda v_j + 1)^2} \sum
_{i \in k^{- 1}(j)} \bar{p}_{b_i}(\alpha, b, \nu)^2,
\]
and we seek zeros of this. The right-hand is clearly an increasing
function of $v_j$, so it is negative somewhere only if it is negative when
$v_j = 0$ where it has the value
%
\begin{equation}
\label{eqdescent-test} \bar{p}_{\nu_j}(\alpha, b, \nu) -
\frac{1}{4} \sum_{i \in k^{- 1}(j)} \bar{p}_{b_i}(
\alpha, b, \nu)^2.
\end{equation}
So that gives us a test for a descent direction: we have a descent
direction if and only if (\ref{eqdescent-test}) is negative.
Conversely, we appear to have $\hat{\nu}_j = 0$ if (\ref{eqdescent-test})
is nonnegative.

\section{\emph{Raphanus sativus} example}

We illustrate the use of this work with three examples, beginning with
a study
of the invasive California wild radish (\emph{Raphanus sativus})
described by \citet{ridley}.
For each individual, three
response variables are observed, connected by the following graphical model:
\[
\begin{CD} 1 @>\mathrm{Ber}>> y_1 @>\mathrm{0\mbox{-}Poi}>>
y_2 @>\mathrm{Poi}>> y_3
\end{CD}
\]
with $y_1$ being an indicator of whether any flowers were produced,
$y_2$ being the count of the number of flowers produced,
$y_3$ being the count of the number of fruits produced,
the unconditional distribution of $y_1$ being Bernoulli,
the conditional distribution of $y_2$ given $y_1$ being zero-truncated Poisson,
and the conditional distribution of $y_3$ given $y_2$ being Poisson.
(The combination of a Bernoulli arrow followed by a zero-truncated Poisson
arrow gives a combined zero-inflated Poisson distribution, that is,
the unconditional distribution of $y_2$ is zero-inflated Poisson.)

These data are found in the data set \texttt{radish} in the R package
\texttt{aster}.
They come from a designed experiment started with seeds collected from
three large wild populations of northern, coastal California wild radish
and three populations of southern, inland California wild radish. Thus, we
have populations nested within region.

Plants were grown at two experimental sites,
one northern, coastal California field site located
at Point Reyes National Seashore and one
southern, inland site located
at the University of California Riverside Agricultural Experiment Station.
Thus, we have blocks nested within site.

The issue of main scientific interest is the interaction of region and
site, which is indicative of local adaptation when the pattern of mean
values shows that each population has higher fitness in its home
environment than
in other environments. Testing significance of this
interaction is complicated by the nesting of populations within region and
blocks within site and the goal of scientists to account for variation due
to these nested factors in evaluating effects of the higher factors.

The best surrogate of fitness in these data is the number of fruits produced.
Thus, we form the ``interaction'' with
the indicator of this component and all scientifically
interesting predictors [see Section~5 of \citet{gws}
or Section~4 of \citet{aster-random}].

The traditional way to deal with a situation like this is to treat the
population effects as random (within region) and the block effects as
random (within site). When we fit this model
[see the technical report \citet{aster-random} for details],
we obtained positive and statistically significantly greater than zero
estimates of both variance components and an estimate 0.499 with
standard error
0.012 for the fixed effect that is the scientifically important site-region
interaction parameter.

\citet{ridley} did not do a random effects aster analysis
because it had not yet been invented. Nevertheless, the conclusions from
their fixed effect aster analysis hold up.
The main conclusion of interest is that there is evidence of
local adaptation. This is indicated by
the statistical significance of the fixed effect for region-site interaction
together with
the pattern of mean values for the different populations in the two
sites, showing that populations growing near to their sampling locations
had higher fitness than in the other location
as found by \citet{ridley}.

The fact that random effects analysis and fixed effects analysis agree
qualitatively on this one example does not, of course, imply that they
would agree on all examples. In these data the region-site interaction
is very large and almost any sensible statistical analysis would show it.
When the interaction is not so large, the analysis done will make a difference.

The analysis reported above is based on the approximations derived in the
theory section. We are using the log approximate integrated likelihood
and its Hessian matrix to do likelihood inference. But what if its
approximations are not valid?
Section~6 of \citet{aster-random} does a parametric bootstrap of this analysis.
It turns out that a 95\% confidence interval for the parameter of interest
(the region-site interaction) does not change much, but other aspects of
the parametric bootstrap are interesting. Sampling distributions of
the estimates of the variance components (as simulated by the parametric
bootstrap) turn out to be highly nonnormal, and these estimators have bias
that is a significant fraction of their standard errors.

\section{\emph{Avena barbata} example}

We use data on the slender wild oat (\emph{Avena barbata})
described by \citet{latta} and contained in the data set \texttt{oats}
in the R contributed package \texttt{aster}. For each individual, two
response variables are observed, connected by the following graphical model:
\[
\begin{CD} 1 @>\mathrm{Ber}>> y_1 @>\mathrm{0\mbox{-}Poi}>>
y_2
\end{CD}
\]
with $y_1$ being an indicator of survival and
$y_2$ being the count of the number of spikelets (compound flowers) produced,
the unconditional distribution of $y_1$ being Bernoulli, and
the conditional distribution of $y_2$ given $y_1$ being zero-truncated Poisson.

These data come from a designed experiment started with seeds collected
in the 1980s in northern California of the xeric (found in drier regions)
and mesic (found in less dry regions) ecotypes.
The variable \texttt{Gen} is the ecotype (``\texttt{X}'' or ``\texttt{M}'').
The variable \texttt{Fam} is the accession (nested within \texttt{Gen}).
The variable \texttt{Site} is the site.
The variable \texttt{Year} is the year (2003 to 2007).
The experimental sites were at the Sierra Foothills Research and Extension
Center (\mbox{\texttt{Site == }``\texttt{SF}''}), which is northeast of Sacramento on the east
side of the Central Valley of California, and at the Hopland Research and
Extension center (\mbox{\texttt{Site = }``\texttt{Hop}''}), which is in the California Coastal
Ranges north of San Francisco. Hopland receives 30\% more rainfall and
has a less severe summer drought than the Sierra foothills.
The best surrogate of fitness in these data is the number of spikelets
produced. Thus, we form the ``interaction'' with
the indicator of this component and all scientifically
interesting predictors.

In the previous analysis [\citet{latta}] a linear mixed model was used,
despite the response being highly nonnormal, because no better tool was
available. Here we reanalyze these data using the same random effects
structure in an aster model.

\begin{table}[h]
\begin{tabular}{@{}lc@{}}
\hline
Effect & \multicolumn{1}{c}{Type} \\
\hline
Site & Fixed \\
Year & Random \\
Gen & Fixed \\
Fam & Random \\
$\mbox{Gen} * \mbox{Site}$ & Fixed \\
$\mbox{Gen} * \mbox{Year}$ & Random \\
$\mbox{Site} * \mbox{Fam}$ & Random \\
$\mbox{Year} * \mbox{Fam}$ & Random\\
\hline
\end{tabular}
\end{table}

We have only three fixed effects parameters because there are only two
levels of \texttt{Site} and two levels of \texttt{Gen}. There are five
variance components, one for each row of the table having random type.

All variance components are estimated to be significantly different
from zero
except for the \texttt{Fam} random effect, which is estimated to be exactly
zero.

The results of the reanalysis agree qualitatively with the original
analysis.
Local adaptation, which would have been shown by a statistically
significant site-ecotype ($\mbox{Gen} * \mbox{Site}$) interaction,
was not found in either analysis (for this interaction,
the aster random effects analysis
obtained the point estimate 0.091 and standard error 0.143).
Moreover, the pattern of mean values was not consistent with local adaptation.
\citet{latta} found that the mesic ecotype had higher fitness (survived
and reproduced better) in all environments. This means that even if
the site-ecotype interaction had been statistically significant, it would
not have indicated local adaptation.

\section{\emph{Chamaecrista fasciculata} data}

We use data on the partridge pea (\emph{Chamaecrista fasciculata})
described by Etterson (\citeyear{etterson-i,etterson-ii}) and \citet{etterson-shaw} and
contained in the data set \texttt{chamae3}
in the R contributed package \texttt{aster}.
\emph{C. fasciculata} grows in the Great Plains of North America
from southern Minnesota to Mexico.
Three focal populations were sampled in the following locations:
\begin{enumerate}
\item Kellog-Weaver Dunes, Wabasha County, Minnesota;
\item Konza Prairie, Riley County, Kansas;
\item Pontotoc Ridge, Pontotoc County, Oklahoma.
\end{enumerate}
These sites are progressively more arid from north to south and also
differ in other characteristics.
Seed pods were collected from 200 plants in each of these three natural
populations. From these, plants were grown and crosses were done;
parent plants are indicated by the variables
\texttt{SIRE} and \texttt{DAM} in the data set.
The
resulting seeds were germinated and established as seedlings in the greenhouse
and then planted using a randomized block
design [\citet{etterson-ii}] in three field sites:
\begin{enumerate}
\item[``\texttt{O}''] Robert S. Kerr Environmental Research Center,
Ada, Oklahoma;
\item[``\texttt{K}''] Konza Prairie Research Natural Area, Manhatten,
Kansas;
\item[``\texttt{M}''] University of Minnesota, St. Paul Minnesota.
\end{enumerate}
The Oklahoma field site was 30 km northwest of the Oklahoma natural population;
the Kansas field site was 5 km from the Kansas natural population;
the Minnesota field site was 110 km northwest of the Minnesota natural
population.
For each individual, two
response variables are observed, connected by the following graphical model:
\[
\begin{CD} 1 @>\mathrm{Ber}>> y_1 @>\mathrm{0\mbox{-}Poi}>>
y_2
\end{CD}
\]
with $y_1$ being an indicator of whether any fruits were produced,
$y_2$ being the count of the number of fruits produced,
the unconditional distribution of $y_1$ being Bernoulli,
and the conditional distribution of $y_2$ given $y_1$ being
zero-truncated Poisson.

We here consider a subset of data previously analyzed by nonaster methods
by Etterson (\citeyear{etterson-i,etterson-ii}) and \citet{etterson-shaw} and by aster
without allowing for random (genetic) effects by \citet{aster2}.
Though seed counts were also observed, the complexity of the
seed count data makes analysis difficult [\citet{aster2}], so it does
not serve
as a good example. Thus, here we analyze only the pod number data,
which does have straightforward aster analysis
and serves as a better example, even though
this makes our reanalysis not really comparable with the analysis
in \citet{etterson-ii} which does use the seed counts.
To aid design of future experiments,
\citet{aster2}, page E43, explain two alternative experimental
designs that permit straightforward aster analysis (including
random effects aster models).
\citet{stanton-geddes-et-al} used one of these designs.

Individuals descended from all three natural populations were planted
in all three field sites, so these data can address local adaptation and
previous analyses [\citet{etterson-ii}, Discussion] did find local adaptation.
But local adaptation is not the main point of interest for our analysis here.
Instead we investigate sire and dam effects, which we treat as random effects,
as did the previous conventional quantitative genetics analysis
[\citet{etterson-ii}].
We focus on sire effects because in this experimental design sire effects
are expected to correspond closely to pure breeding values (additive
genetic effects) but dam effects
confound additive with maternal and dominance effects.

Because the biology that leads to fitness may differ at different sites
and in different populations, we did nine separate analyses, one for each
population-site combination.

We found that the sire variance components for the Minnesota and Oklahoma
natural population are not close to statistically significant at the
Minnesota field site. All the other sire variance components are at
least borderline statistically significant.

Our analysis produces not only estimates of the variance components but also
estimates of the random effects (these are the penalized quasi-likelihood
estimates $b^*$ described in the theory sections).
As a matter of purely statistical interest, we examined the Gaussianity of
the random effects. They seemed to be normal (or at least not statistically
significantly nonnormal by a Shapiro--Wilk test). We conjectured that this
apparent normality was due to the estimation procedure, but this turned out
not to be the case, since when we redid penalized quasi-likelihood estimates
of the random effects using much smaller penalties than the maximum likelihood
penalty, the random effects still seemed normal.

\begin{figure}

\includegraphics{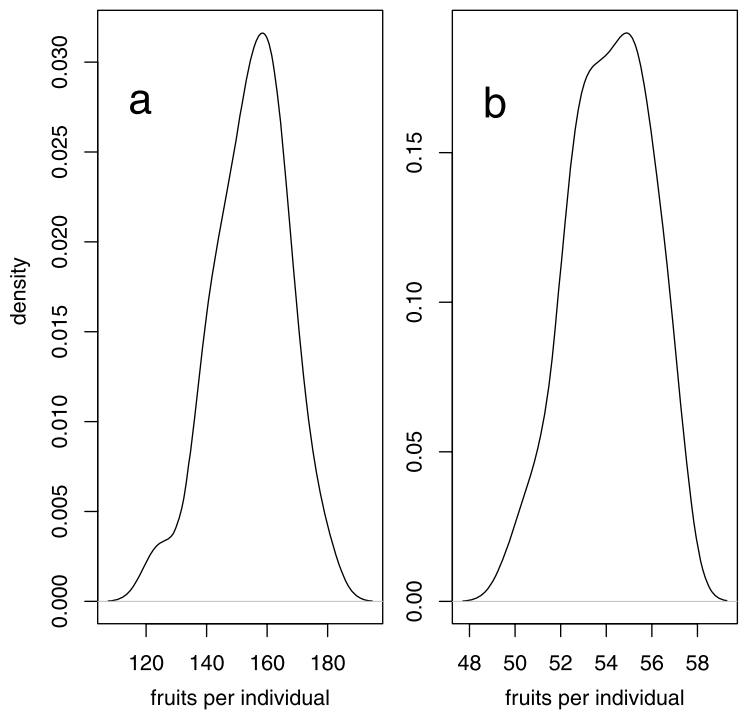}

\caption{Density plot of sire effects on the mean value parameter
scale for
an individual in block 1 having the various sire effects in the data.
Panel \textup{(a)} is the Kansas population in the Kansas field site.
Panel \textup{(b)} is the Kansas population in the Oklahoma field site.}
\label{figone}
\end{figure}

For interpretability, biologists want random effects mapped to the mean value
parameter scale (rather than the canonical parameter scale where they
originally are). To illustrate this, we mapped the sire effects for two
population-site pairs to the mean value parameter scale, setting the dam
effects to be zero (the middling value) and setting the block effect to be
block~1 (each site was divided into blocks). Figure~\ref{figone} shows
these plots; for details of how they were done see Section~8.6 of
\citet{aster-random}.
%
This figure was made using the default smoothing parameter selection
of the R function \texttt{density}. The apparent non-Gaussianity is
not statistically significant
[Shapiro--Wilk test, \citet{aster-random}, Section~8.6].

Thus, the aster model can include random effects for parents and
permit quantitative genetic inference for fitness variation.

\section{Discussion}

Our methods are founded on two approximations taken from \citet{b+c}.
Our technical innovations are that we provide derivatives of the log approximate
integrated likelihood (Section~\ref{secLAIL})
and the test (\ref{eqdescent-test}) for when variance components are zero,
which is
based on the theory of constrained optimization.
Our methods work well when there are multiple variance components.
Two examples had two variance components and one had five variance components.
However, problems arise when there are thousands of random effects, and
especially when there is one random effect per
individual. Since quantitative genetics traditionally does have individual
random effects as well as parental random effects, our methods are not
fully comparable to traditional quantitative genetics.

\citet{rutter-et-al} have already used aster models with random effects
for an analysis of the effect of known spontaneous mutations on fitness in
\emph{Arabidopsis thaliana} grown in different environments.

Past experience [\citet{sung}] with examples taken from the literature shows
that log integrated likelihoods are often far from quadratic,
even when no approximations are done, in which case
asymptotics based on Fisher information are inaccurate.
Thus, we recommend the parametric bootstrap here, as we do whenever
there is
doubt about the validity of asymptotics for parametric inference.
We illustrate the parametric bootstrap for one of our examples.
This need for doing a parametric bootstrap is another reason for preferring
computationally efficient methods. In particular, it is incredibly time
consuming to bootstrap Monte Carlo calculations if Monte Carlo run lengths
are long enough for accurate calculation.

\citet{b+c} introduced yet another approximation that is supposed to
be analogous to restricted maximum likelihood (REML), but we did not
use this.
First, the analogy to REML is weak, and this method has no provable
mathematical properties. Second, we do not see how this method extends to
general exponential family models, such as aster models. Third, even in
conventional linear mixed models, REML does not seem to be appropriate when
the parameters of interest are fixed effects, which is often the case in
biology and is the case in some of our examples.




\printaddresses


\begin{thebibliography}{37}

\bibitem[\protect\citeauthoryear{Anderson}{2003}]{anderson}
\begin{bbook}[mr]
\bauthor{\bsnm{Anderson},~\bfnm{T.~W.}\binits{T.~W.}}
(\byear{2003}).
\btitle{An Introduction to Multivariate Statistical Analysis},
\bedition{3rd} ed.
\bpublisher{Wiley}, \blocation{Hoboken, NJ}.
\bid{mr={1990662}}
\bptok{imsref}%
\end{bbook}
\endbibitem

\bibitem[\protect\citeauthoryear{Barndorff-Nielsen}{1978}]{barndorff}
\begin{bbook}[author]
\bauthor{\bsnm{Barndorff-Nielsen},~\bfnm{O.}\binits{O.}}
(\byear{1978}).
\btitle{Information and Exponential Families}.
\bpublisher{Wiley}, \blocation{Chichester}.
\bid{mr={0489333}}
\bptok{imsref}%
\end{bbook}
\endbibitem

\bibitem[\protect\citeauthoryear{Booth and Hobert}{1999}]{booth-hobert}
\begin{barticle}[author]
\bauthor{\bsnm{Booth},~\bfnm{J.}\binits{J.}} \AND
  \bauthor{\bsnm{Hobert},~\bfnm{J.~P.}\binits{J.~P.}}
(\byear{1999}).
\btitle{Maximizing generalized linear mixed model likelihoods with an automated
  {M}onte {C}arlo {EM} algorithm}.
\bjournal{J. R. Stat. Soc. Ser. B Stat. Methodol.}
\bvolume{61}
\bpages{265--285}.
\bptok{imsref}%
\end{barticle}
\endbibitem

\bibitem[\protect\citeauthoryear{Breslow and Clayton}{1993}]{b+c}
\begin{barticle}[author]
\bauthor{\bsnm{Breslow},~\bfnm{N.~E.}\binits{N.~E.}} \AND
  \bauthor{\bsnm{Clayton},~\bfnm{D.~G.}\binits{D.~G.}}
(\byear{1993}).
\btitle{Approximate inference in generalized linear mixed models}.
\bjournal{J. Amer. Statist. Assoc.}
\bvolume{88}
\bpages{9--25}.
\bptok{imsref}%
\end{barticle}
\endbibitem

\bibitem[\protect\citeauthoryear{Crouch and
  Spiegelman}{1990}]{crouch-spiegelman}
\begin{barticle}[mr]
\bauthor{\bsnm{Crouch},~\bfnm{Edmund A.~C.}\binits{E.~A.~C.}} \AND
  \bauthor{\bsnm{Spiegelman},~\bfnm{Donna}\binits{D.}}
(\byear{1990}).
\btitle{The evaluation of integrals of the form {$\int\sp {+\infty}\sb
  {-\infty}f(t)\exp(-t\sp 2)\,dt$}: Application to logistic-normal models}.
\bjournal{J. Amer. Statist. Assoc.}
\bvolume{85}
\bpages{464--469}.
\bid{issn={0162-1459}, mr={1141749}}
\bptok{imsref}%
\end{barticle}
\endbibitem

\bibitem[\protect\citeauthoryear{Etterson}{2004a}]{etterson-i}
\begin{barticle}[pbm]
\bauthor{\bsnm{Etterson},~\bfnm{Julie~R.}\binits{J.~R.}}
(\byear{2004}a).
\btitle{Evolutionary potential of Chamaecrista fasciculata in relation to
  climate change. I. Clinal patterns of selection along an environmental
  gradient in the great plains}.
\bjournal{Evolution}
\bvolume{58}
\bpages{1446--1458}.
\bid{issn={0014-3820}, pmid={15341148}}
\bptok{imsref}%
\end{barticle}
\endbibitem

\bibitem[\protect\citeauthoryear{Etterson}{2004b}]{etterson-ii}
\begin{barticle}[pbm]
\bauthor{\bsnm{Etterson},~\bfnm{Julie~R.}\binits{J.~R.}}
(\byear{2004}b).
\btitle{Evolutionary potential of Chamaecrista fasciculata in relation to
  climate change. II. Genetic architecture of three populations reciprocally
  planted along an environmental gradient in the great plains}.
\bjournal{Evolution}
\bvolume{58}
\bpages{1459--1471}.
\bid{issn={0014-3820}, pmid={15341149}}
\bptok{imsref}%
\end{barticle}
\endbibitem

\bibitem[\protect\citeauthoryear{Etterson and Shaw}{2001}]{etterson-shaw}
\begin{barticle}[pbm]
\bauthor{\bsnm{Etterson},~\bfnm{J.~R.}\binits{J.~R.}} \AND
  \bauthor{\bsnm{Shaw},~\bfnm{R.~G.}\binits{R.~G.}}
(\byear{2001}).
\btitle{Constraint to adaptive evolution in response to global warming}.
\bjournal{Science}
\bvolume{294}
\bpages{151--154}.
\bid{doi={10.1126/science.1063656}, issn={0036-8075}, pii={294/5540/151},
  pmid={11588260}}
\bptok{imsref}%
\end{barticle}
\endbibitem

\bibitem[\protect\citeauthoryear{Fisher}{1918}]{fisher18}
\begin{barticle}[author]
\bauthor{\bsnm{Fisher},~\bfnm{Ronald~Alymer}\binits{R.~A.}}
(\byear{1918}).
\btitle{The correlation between relatives on the supposition of {Mendelian}
  inheritance}.
\bjournal{Trans. R. Soc. Edinburgh}
\bvolume{52}
\bpages{399--433}.
\bptok{imsref}%
\end{barticle}
\endbibitem

\bibitem[\protect\citeauthoryear{Fletcher}{1987}]{fletcher}
\begin{bbook}[mr]
\bauthor{\bsnm{Fletcher},~\bfnm{R.}\binits{R.}}
(\byear{1987}).
\btitle{Practical Methods of Optimization},
\bedition{2nd} ed.
\bpublisher{Wiley}, \blocation{Chichester}.
\bid{mr={0955799}}
\bptok{imsref}%
\end{bbook}
\endbibitem

\bibitem[\protect\citeauthoryear{Geyer}{1994}]{mcmcml}
\begin{barticle}[mr]
\bauthor{\bsnm{Geyer},~\bfnm{Charles~J.}\binits{C.~J.}}
(\byear{1994}).
\btitle{On the convergence of {M}onte {C}arlo maximum likelihood calculations}.
\bjournal{J. R. Stat. Soc. Ser. B Stat. Methodol.}
\bvolume{56}
\bpages{261--274}.
\bid{issn={0035-9246}, mr={1257812}}
\bptok{imsref}%
\end{barticle}
\endbibitem

\bibitem[\protect\citeauthoryear{Geyer}{2009}]{gdor}
\begin{barticle}[mr]
\bauthor{\bsnm{Geyer},~\bfnm{Charles~J.}\binits{C.~J.}}
(\byear{2009}).
\btitle{Likelihood inference in exponential families and directions of
  recession}.
\bjournal{Electron. J. Stat.}
\bvolume{3}
\bpages{259--289}.
\bid{doi={10.1214/08-EJS349}, issn={1935-7524}, mr={2495839}}
\bptok{imsref}%
\end{barticle}
\endbibitem

\bibitem[\protect\citeauthoryear{Geyer}{2012}]{package-aster}
\begin{bmisc}[author]
\bauthor{\bsnm{Geyer},~\bfnm{C.~J.}\binits{C.~J.}}
(\byear{2012}).
\bhowpublished{R package aster, version 0.8-19.
Available at \texttt{\href{http://cran.r-project.org/package=aster}{http://cran.}
\href{http://cran.r-project.org/package=aster}{r-project.org/package=aster}}}.
\bptok{imsref}%
\end{bmisc}
\endbibitem

\bibitem[\protect\citeauthoryear{Geyer}{2013}]{simple}
\begin{bincollection}[author]
\bauthor{\bsnm{Geyer},~\bfnm{C.~J.}\binits{C.~J.}}
(\byear{2013}).
\btitle{Asymptotics of maximum likelihood without the {LLN} or {CLT} or sample
  size going to infinity}.
In \bbooktitle{Multivariate Statistics in Modern Statistical Analysis: A~Festschrift for {M}orris {L}. {E}aton}
(\beditor{\bfnm{G.}\binits{G.}~\bsnm{Jones}} \AND
  \beditor{\bfnm{X.}\binits{X.}~\bsnm{Shen}}, eds.)
\bvolume{10}
\bpages{1--24}.
\bpublisher{IMS}, \blocation{Hayward, CA}.
\bptok{imsref}%
\end{bincollection}
\endbibitem

\bibitem[\protect\citeauthoryear{Geyer and Thompson}{1992}]{g+t}
\begin{barticle}[mr]
\bauthor{\bsnm{Geyer},~\bfnm{Charles~J.}\binits{C.~J.}} \AND
  \bauthor{\bsnm{Thompson},~\bfnm{Elizabeth~A.}\binits{E.~A.}}
(\byear{1992}).
\btitle{Constrained {M}onte {C}arlo maximum likelihood for dependent data}.
\bjournal{J. R. Stat. Soc. Ser. B Stat. Methodol.}
\bvolume{54}
\bpages{657--699}.
\bid{issn={0035-9246}, mr={1185217}}
\bptnote{check related}%
\bptok{imsref}%
\end{barticle}
\endbibitem

\bibitem[\protect\citeauthoryear{Geyer, Wagenius and Shaw}{2007}]{gws}
\begin{barticle}[mr]
\bauthor{\bsnm{Geyer},~\bfnm{Charles~J.}\binits{C.~J.}},
  \bauthor{\bsnm{Wagenius},~\bfnm{Stuart}\binits{S.}} \AND
  \bauthor{\bsnm{Shaw},~\bfnm{Ruth~G.}\binits{R.~G.}}
(\byear{2007}).
\btitle{Aster models for life history analysis}.
\bjournal{Biometrika}
\bvolume{94}
\bpages{415--426}.
\bid{doi={10.1093/biomet/asm030}, issn={0006-3444}, mr={2380569}}
\bptok{imsref}%
\end{barticle}
\endbibitem

\bibitem[\protect\citeauthoryear{Geyer et~al.}{2012}]{aster-random}
\begin{bmisc}[author]
\bauthor{\bsnm{Geyer},~\bfnm{Charles~J.}\binits{C.~J.}},
  \bauthor{\bsnm{Ridley},~\bfnm{Caroline~E.}\binits{C.~E.}},
  \bauthor{\bsnm{Latta},~\bfnm{Robert~G.}\binits{R.~G.}},
  \bauthor{\bsnm{Etterson},~\bfnm{Julie~R.}\binits{J.~R.}} \AND
  \bauthor{\bsnm{Shaw},~\bfnm{Ruth~G.}\binits{R.~G.}}
(\byear{2012}).
\bhowpublished{Aster models with random effects via penalized
likelihood.
Technical Report 692,
Univ. Minnesota School of Statistics. Available at
\url{http://purl.umn.edu/135870}}.
\bptok{imsref}%
\end{bmisc}
\endbibitem

\bibitem[\protect\citeauthoryear{Hummel, Hunter and
  Handcock}{2012}]{hummel-et-al}
\begin{barticle}[mr]
\bauthor{\bsnm{Hummel},~\bfnm{Ruth~M.}\binits{R.~M.}},
  \bauthor{\bsnm{Hunter},~\bfnm{David~R.}\binits{D.~R.}} \AND
  \bauthor{\bsnm{Handcock},~\bfnm{Mark~S.}\binits{M.~S.}}
(\byear{2012}).
\btitle{Improving simulation-based algorithms for fitting {ERGM}s}.
\bjournal{J. Comput. Graph. Statist.}
\bvolume{21}
\bpages{920--939}.
\bid{doi={10.1080/10618600.2012.679224}, issn={1061-8600}, mr={3005804}}
\bptok{imsref}%
\end{barticle}
\endbibitem

\bibitem[\protect\citeauthoryear{Hunter et~al.}{2008}]{hunter-et-al}
\begin{barticle}[pbm]
\bauthor{\bsnm{Hunter},~\bfnm{David~R.}\binits{D.~R.}},
  \bauthor{\bsnm{Handcock},~\bfnm{Mark~S.}\binits{M.~S.}},
  \bauthor{\bsnm{Butts},~\bfnm{Carter~T.}\binits{C.~T.}},
  \bauthor{\bsnm{Goodreau},~\bfnm{Steven~M.}\binits{S.~M.}} \AND
  \bauthor{\bsnm{Morris},~\bfnm{Martina}\binits{M.}}
(\byear{2008}).
\btitle{ergm: A Package to fit, simulate and diagnose exponential-family models
  for networks}.
\bjournal{J. Stat. Softw.}
\bvolume{24}
\bpages{nihpa54860}.
\bid{issn={1548-7660}, mid={NIHMS54860}, pmcid={2743438}, pmid={19756229}}
\bptok{imsref}%
\end{barticle}
\endbibitem

\bibitem[\protect\citeauthoryear{Latta}{2009}]{latta}
\begin{barticle}[author]
\bauthor{\bsnm{Latta},~\bfnm{R.~G.}\binits{R.~G.}}
(\byear{2009}).
\btitle{Testing for local adaptation in \textit{Avena barbata}, a classic
  example of ecotypic divergence}.
\bjournal{Molecular Ecology}
\bvolume{18}
\bpages{3781--3791}.
\bptok{imsref}%
\end{barticle}
\endbibitem

\bibitem[\protect\citeauthoryear{Le~Cam and Yang}{2000}]{lecam-yang}
\begin{bbook}[mr]
\bauthor{\bsnm{Le~Cam},~\bfnm{Lucien}\binits{L.}} \AND
  \bauthor{\bsnm{Yang},~\bfnm{Grace~Lo}\binits{G.~L.}}
(\byear{2000}).
\btitle{Asymptotics in Statistics: Some Basic Concepts},
\bedition{2nd} ed.
\bpublisher{Springer}, \blocation{New York}.
\bid{doi={10.1007/978-1-4612-1166-2}, mr={1784901}}
\bptok{imsref}%
\end{bbook}
\endbibitem

\bibitem[\protect\citeauthoryear{McCulloch}{2003}]{mccullough}
\begin{bbook}[mr]
\bauthor{\bsnm{McCulloch},~\bfnm{Charles~E.}\binits{C.~E.}}
(\byear{2003}).
\btitle{Generalized Linear Mixed Models}.
\bseries{NSF-CBMS Regional Conference Series in Probability and Statistics}
\bvolume{7}.
\bpublisher{IMS}, \blocation{Beachwood, OH}.
\bid{mr={1993816}}
\bptok{imsref}%
\end{bbook}
\endbibitem

\bibitem[\protect\citeauthoryear{Nocedal and Wright}{1999}]{nocedal-wright}
\begin{bbook}[mr]
\bauthor{\bsnm{Nocedal},~\bfnm{Jorge}\binits{J.}} \AND
  \bauthor{\bsnm{Wright},~\bfnm{Stephen~J.}\binits{S.~J.}}
(\byear{1999}).
\btitle{Numerical Optimization}.
\bpublisher{Springer}, \blocation{New York}.
\bid{doi={10.1007/b98874}, mr={1713114}}
\bptok{imsref}%
\end{bbook}
\endbibitem

\bibitem[\protect\citeauthoryear{Okabayashi and Geyer}{2011}]{okabayashi-geyer}
\begin{barticle}[author]
\bauthor{\bsnm{Okabayashi},~\bfnm{S.}\binits{S.}} \AND
  \bauthor{\bsnm{Geyer},~\bfnm{C.~J.}\binits{C.~J.}}
(\byear{2011}).
\btitle{Gradient-based search for maximum likelihood in exponential families}.
\bjournal{Electron. J. Stat.}
\bvolume{6}
\bpages{123--147}.
\bid{mr={2879674}}
\bptok{imsref}%
\end{barticle}
\endbibitem

\bibitem[\protect\citeauthoryear{Penttinen}{1984}]{penttinen}
\begin{bmisc}[author]
\bauthor{\bsnm{Penttinen},~\bfnm{A.}\binits{A.}}
(\byear{1984}).
\bhowpublished{Modelling interations in spatial point patterns: Parameter estimation
  by the maximum likelihood method.
Jyv\"{a}skyl\"{a} Studies in Computer Science, Economics and Statistics
  No. 7,
Univ. Jyv\"{a}skyl\"{a}}.
\bptok{imsref}%
\end{bmisc}
\endbibitem

\bibitem[\protect\citeauthoryear{Ridley and Ellstrand}{2010}]{ridley}
\begin{barticle}[author]
\bauthor{\bsnm{Ridley},~\bfnm{C.~E.}\binits{C.~E.}} \AND
  \bauthor{\bsnm{Ellstrand},~\bfnm{N.~C.}\binits{N.~C.}}
(\byear{2010}).
\btitle{Rapid evolution of morphology and adaptive life history in the invasive
  California wild radish (\textit{Raphanus sativus}) and the implications for
  management}.
\bjournal{Evolutionary Applications}
\bvolume{3}
\bpages{64--76}.
\bptok{imsref}%
\end{barticle}
\endbibitem

\bibitem[\protect\citeauthoryear{Rockafellar and Wets}{2004}]{raw}
\begin{bbook}[author]
\bauthor{\bsnm{Rockafellar},~\bfnm{R.~T.}\binits{R.~T.}} \AND
  \bauthor{\bsnm{Wets},~\bfnm{R.~J.~B.}\binits{R.~J.~B.}}
(\byear{2004}).
\btitle{Variational Analysis},
\bedition{2nd} corrected printing.
\bpublisher{Springer}, \blocation{Berlin}.
\bid{mr={1491362}}
\bptok{imsref}%
\end{bbook}
\endbibitem

\bibitem[\protect\citeauthoryear{Rutter et~al.}{2012}]{rutter-et-al}
\begin{barticle}[pbm]
\bauthor{\bsnm{Rutter},~\bfnm{Matthew~T.}\binits{M.~T.}},
  \bauthor{\bsnm{Roles},~\bfnm{Angela}\binits{A.}},
  \bauthor{\bsnm{Conner},~\bfnm{Jeffrey~K.}\binits{J.~K.}},
  \bauthor{\bsnm{Shaw},~\bfnm{Ruth~G.}\binits{R.~G.}},
  \bauthor{\bsnm{Shaw},~\bfnm{Frank~H.}\binits{F.~H.}},
  \bauthor{\bsnm{Schneeberger},~\bfnm{Korbinian}\binits{K.}},
  \bauthor{\bsnm{Ossowski},~\bfnm{Stephan}\binits{S.}},
  \bauthor{\bsnm{Weigel},~\bfnm{Detlef}\binits{D.}} \AND
  \bauthor{\bsnm{Fenster},~\bfnm{Charles~B.}\binits{C.~B.}}
(\byear{2012}).
\btitle{Fitness of Arabidopsis thaliana mutation accumulation lines whose
  spontaneous mutations are known}.
\bjournal{Evolution}
\bvolume{66}
\bpages{2335--2339}.
\bid{doi={10.1111/j.1558-5646.2012.01583.x}, issn={1558-5646}, pmid={22759306}}
\bptok{imsref}%
\end{barticle}
\endbibitem

\bibitem[\protect\citeauthoryear{Shaw, Geyer and Shaw}{2002}]{shaw-geyer-shaw}
\begin{barticle}[pbm]
\bauthor{\bsnm{Shaw},~\bfnm{Frank~H.}\binits{F.~H.}},
  \bauthor{\bsnm{Geyer},~\bfnm{Charles~J.}\binits{C.~J.}} \AND
  \bauthor{\bsnm{Shaw},~\bfnm{Ruth~G.}\binits{R.~G.}}
(\byear{2002}).
\btitle{A comprehensive model of mutations affecting fitness and inferences for
  Arabidopsis thaliana}.
\bjournal{Evolution}
\bvolume{56}
\bpages{453--463}.
\bid{issn={0014-3820}, pmid={11989677}}
\bptok{imsref}%
\end{barticle}
\endbibitem

\bibitem[\protect\citeauthoryear{Shaw et~al.}{1999}]{promislow}
\begin{barticle}[author]
\bauthor{\bsnm{Shaw},~\bfnm{F.~H.}\binits{F.~H.}},
  \bauthor{\bsnm{Promislow},~\bfnm{D.~E.~L.}\binits{D.~E.~L.}},
  \bauthor{\bsnm{Tatar},~\bfnm{M.}\binits{M.}},
  \bauthor{\bsnm{Hughes},~\bfnm{K.~A.}\binits{K.~A.}} \AND
  \bauthor{\bsnm{Geyer},~\bfnm{C.~J.}\binits{C.~J.}}
(\byear{1999}).
\btitle{Towards reconciling inferences concerning genetic variation in
  senescence}.
\bjournal{Genetics}
\bvolume{152}
\bpages{553--566}.
\bptok{imsref}%
\end{barticle}
\endbibitem

\bibitem[\protect\citeauthoryear{Shaw et~al.}{2008}]{aster2}
\begin{barticle}[pbm]
\bauthor{\bsnm{Shaw},~\bfnm{Ruth~G.}\binits{R.~G.}},
  \bauthor{\bsnm{Geyer},~\bfnm{Charles~J.}\binits{C.~J.}},
  \bauthor{\bsnm{Wagenius},~\bfnm{Stuart}\binits{S.}},
  \bauthor{\bsnm{Hangelbroek},~\bfnm{Helen~H.}\binits{H.~H.}} \AND
  \bauthor{\bsnm{Etterson},~\bfnm{Julie~R.}\binits{J.~R.}}
(\byear{2008}).
\btitle{Unifying life-history analyses for inference of fitness and population
  growth}.
\bjournal{Am. Nat.}
\bvolume{172}
\bpages{E35--E47}.
\bid{doi={10.1086/588063}, issn={1537-5323}, pmid={18500940}}
\bptok{imsref}%
\end{barticle}
\endbibitem

\bibitem[\protect\citeauthoryear{Stanton-Geddes, Shaw and
  Tiffin}{2012}]{stanton-geddes-et-al}
\begin{barticle}[pbm]
\bauthor{\bsnm{Stanton-Geddes},~\bfnm{John}\binits{J.}},
  \bauthor{\bsnm{Shaw},~\bfnm{Ruth~G.}\binits{R.~G.}} \AND
  \bauthor{\bsnm{Tiffin},~\bfnm{Peter}\binits{P.}}
(\byear{2012}).
\btitle{Interactions between soil habitat and geographic range location affect
  plant fitness}.
\bjournal{PLoS ONE}
\bvolume{7}
\bpages{e36015}.
\bid{doi={10.1371/journal.pone.0036015}, issn={1932-6203},
  pii={PONE-D-12-04401}, pmcid={3355151}, pmid={22615745}}
\bptok{imsref}%
\end{barticle}
\endbibitem

\bibitem[\protect\citeauthoryear{Sung and Geyer}{2007}]{sung}
\begin{barticle}[mr]
\bauthor{\bsnm{Sung},~\bfnm{Yun~Ju}\binits{Y.~J.}} \AND
  \bauthor{\bsnm{Geyer},~\bfnm{Charles~J.}\binits{C.~J.}}
(\byear{2007}).
\btitle{Monte {C}arlo likelihood inference for missing data models}.
\bjournal{Ann. Statist.}
\bvolume{35}
\bpages{990--1011}.
\bid{doi={10.1214/009053606000001389}, issn={0090-5364}, mr={2341695}}
\bptok{imsref}%
\end{barticle}
\endbibitem

\bibitem[\protect\citeauthoryear{R Core Team}{2012}]{rcore}
\begin{bmisc}[author]
\borganization{R Core Team}
(\byear{2012}).
\bhowpublished{R: A language and environment for statistical computing. R
  foundation for statistical computing, Vienna, Austria. Available at
  \url{http://www.R-project.org/}}.
\bptok{imsref}%
\end{bmisc}
\endbibitem

\bibitem[\protect\citeauthoryear{Thompson and Guo}{1991}]{thompson}
\begin{barticle}[pbm]
\bauthor{\bsnm{Thompson},~\bfnm{E.~A.}\binits{E.~A.}} \AND
  \bauthor{\bsnm{Guo},~\bfnm{S.~W.}\binits{S.~W.}}
(\byear{1991}).
\btitle{Evaluation of likelihood ratios for complex genetic models}.
\bjournal{IMA J. Math. Appl. Med. Biol.}
\bvolume{8}
\bpages{149--169}.
\bid{issn={0265-0746}, pmid={1823088}}
\bptok{imsref}%
\end{barticle}
\endbibitem

\bibitem[\protect\citeauthoryear{Travisano and Shaw}{2013}]{travisano-shaw}
\begin{barticle}[pbm]
\bauthor{\bsnm{Travisano},~\bfnm{Michael}\binits{M.}} \AND
  \bauthor{\bsnm{Shaw},~\bfnm{Ruth~G.}\binits{R.~G.}}
(\byear{2013}).
\btitle{Lost in the map}.
\bjournal{Evolution}
\bvolume{67}
\bpages{305--314}.
\bid{doi={10.1111/j.1558-5646.2012.01802.x}, issn={1558-5646}, pmid={23356605}}
\bptok{imsref}%
\end{barticle}
\endbibitem

\end{thebibliography}
\end{document}